\documentclass[onecollarge,natbib]{svjour2}
\bibpunct{[}{]}{;}{n}{}{,} 
\smartqed  
\usepackage{graphicx}
\usepackage{epsfig}
\usepackage{amsmath,amssymb,amsfonts}
\usepackage{hyperref}
\usepackage{mathrsfs}
\usepackage{bbm}
\usepackage{slashed}
\usepackage{graphicx}
\usepackage{verbatim}

\usepackage[usenames]{color}

\newcommand{\be}{\begin{equation}}
\newcommand{\ee}{\end{equation}}
\newcommand{\bw}{\begin{widetext}}
\newcommand{\ew}{\end{widetext}}
\newcommand{\bi}{\begin{itemize}}
\newcommand{\ei}{\end{itemize}}
\newcommand{\bea}{\begin{eqnarray}}
\newcommand{\eea}{\end{eqnarray}}
\newcommand{\bra}[1]{\langle\,#1\,|}          
\newcommand{\ket}[1]{|\,#1\,\rangle}          
\newcommand{\ud}{\mathrm{d}}

\newcommand{\LCm}{{\scriptscriptstyle -}} 
\newcommand{\LCp}{{\scriptscriptstyle +}}
\newcommand{\LCpm}{{\scriptscriptstyle \pm}}

\newcommand{\LCperp}{{\scriptscriptstyle \perp}}

\newcommand{\tint}[1]{\int\!\ud\tilde{#1}}

\journalname{Few-Body Systems}
\begin{document}

\title{Ellipticity induced in vacuum birefringence
}


\author{Greger Torgrimsson  
}


\institute{G. Torgrimsson \at
              Department of Applied Physics, Chalmers University of Technology, SE-41296 Gothenburg, Sweden \\
              \email{greger.torgrimsson@chalmers.se}           
}

\date{Received: date / Accepted: date}

\maketitle

\begin{abstract}
We consider signals of photon-photon scattering in laser-based, low energy experiments. In particular, we consider the ellipticity induced on a probe beam by a strong background field, and compare it with a recent worldline expression for the photon polarisation flip amplitude. When the probe and the background are plane waves, the ellipticity is equal to the flip amplitude. Here we investigate the ellipticity-amplitude relation for more physical fields.   

\end{abstract}

\section{Introduction}\label{intro}

Strong field QED currently attracts much interest, see \cite{DiPiazza:2011tq} for a review. In a few years, long-awaited experiments will be performed at the next generation of high intensity laser facilities. One of the most interesting processes to study is vacuum birefringence, where a strong background field changes the polarisation of a probe laser beam passing through it. In particular, an initially linearly polarised probe will, after interacting with the background, emerge with elliptical polarisation. The word "vacuum" here distinguishes this process from the analogous process in optics, where birefringence is produced by e.g. a crystal. Vacuum birefringence is not a property of the vacuum, but rather a manifestation of photon-photon scattering. Vacuum birefringence was first predicted by Toll in his PhD thesis \cite{Toll:1952} and has since given rise to a vast literature, see \cite{DiPiazza:2011tq,Dittrich:2000zu,biref1,biref2,Heinzl:2006xc} and references therein. In \cite{Heinzl:2006xc} an experiment was proposed to search for vacuum birefringence using two colliding laser beams. This experiment is now planned to be performed at HIBEF \cite{HIBEF,HP}, and will, if successful, give the first experimental verification of vacuum birefringence along with photon-photon scattering (with only real photons, c.f. \cite{Jarlskog:1974tx,Schumacher:1975kv}).  

In \cite{biref1} we described the two colliding laser beams by (pulsed) plane waves, which is a common first approximation that allows for an exact treatment of the background. The change in the polarisation of the probe was obtained from the expectation value of the electric field operator, and the close relation with the amplitude for a single photon to flip polarisation was emphasised. This calculation was performed in lightfront quantisation, which, as noted in \cite{Neville:1971uc,biref1,Bakker:2013cea,Zhao:2013cma,Zhao:2013jia}, is particularly natural for many strong field problems. The resulting birefringence signal in the low energy limit agrees with known literature results; another way to view this is as (to the best of our knowledge) the first lightfront quantisation derivation of the coefficients in the Euler-Heisenberg effective action. Our results actually go beyond Euler-Heisenberg and so allow us to consider interesting high energy effects. Upcoming experiments, though, will be in the low energy regime. In \cite{biref2} we therefore started with the low energy effective action and derived an expression for the polarisation flip amplitude that allowed us to consider realistic backgrounds and the impact of various collision parameters.

Here we will complement \cite{biref1,biref2} by deriving the ellipticity in the low energy regime. Under certain conditions, the ellipticity is equal to the flip amplitude found in \cite{biref2}, thus generalising the ellipticity-amplitude relation in \cite{biref1}. These proceedings are organised as follows. In Sect.~\ref{Birefringence from lightfront} we will review the derivation of vacuum birefringence in plane waves using lightfront quantisation \cite{biref1}. We will see that, in the low energy regime, the polarisation flip amplitude and the ellipticity are given by a single lightfront time integral, which can be interpreted as an integral over a photon worldline. In this line-integral form the result can be generalised by simply replacing the plane wave background with an arbitrary field. In Sect.~\ref{Ellipticity-section} the ellipticity is derived and the conditions under which it is equal to the flip amplitude are investigated. Sect.~\ref{Probability for polarisation-flip} gives an alternative derivation of the flip amplitude, which is closer to the ellipticity derivation.

\section{Birefringence in lightfront quantisation}\label{Birefringence from lightfront}

In this section we will review the lightfront method to obtain birefringence \cite{biref1} and the worldline integral expression. For reviews of lightfront quantisation see \cite{Brodsky:1997de,Heinzl:1998kz}. Lightfront coordinates are defined by $x^\pm=2x_\mp=x^0\pm x^3$ and $x^\LCperp=\{x^1,x^2\}$. In this section the background $f$ is a pulsed plane wave and we choose coordinates so that it depends on lightfront time $ef_{\mu\nu}=k_{[\mu}a'_{\nu]}(kx)$, with $kx\propto x^\LCp$, $a_{\LCpm}=0$ and $a_\LCperp(\pm\infty)=0$. As noted in \cite{Neville:1971uc} lightfront quantisation combined with the Furry picture is particularly convenient when dealing with plane waves. In the Furry picture the gauge field $A_\mu$ has the same form as in the interaction picture, while the fermion field $\Psi$ is written in terms of Volkov solutions $\psi$ as
\be
\Psi=\tint{p}\;b_{sp}\psi_p u_{sp}+d_{sp}^\dagger \psi_{-p}v_{sp} \qquad \psi=\Big(1+\frac{\slashed{k}\slashed{a}}{2kp}\Big)\exp-i\Big(px+\int\limits^{kx}\frac{2pa-a^2}{2kp}\Big) \;,
\ee
where $\ud\tilde{p}$ is the on-shell Lorentz-invariant measure and the modes satisfy the usual anti-commutation relations. The probe is represented by a coherent state $\ket{P}$ describing a monochromatic plane wave with wave vector $l_\mu$ and linear polarisation $\epsilon_\mu$. The change in polarisation, caused by the interaction with the background, is found by evolving $\ket{P}$ in $x^\LCp$ using the lightfront Hamiltonian $P_\LCp$, and calculating the expectation value of the field operator. $P_\LCp$ has two terms that are quadratic in $e$, but here only the first order term contributes,
\be
H_1=\frac{e}{2}\int\ud x^\LCm\ud x^\LCperp\bar{\Psi}\slashed{A}\Psi \;.
\ee
The expectation values are calculated to $\mathcal{O}(\alpha)$ by first commuting away all mode operators. One finds that one can perform all integrals except for those over lightfront time \cite{biref1}. Defining the single photon amplitude by 
\be
\bra{l',\epsilon'}S\ket{l,\epsilon}=\tilde{\delta}(l,l')(-\epsilon\epsilon'+iT_{\epsilon'\epsilon}) \;,
\ee
where $S$ is the time evolution operator, we find that the probe electric field in the final state, projected onto some vector $\epsilon'$, can be written
\be\label{delta-T-plane-wave}
\epsilon'\langle E\rangle=\text{Re }(-\epsilon'\epsilon+iT_{\epsilon'\epsilon})P_0e^{-ilx}\;,
\ee   
where $P_0$ is a constant and the initial field is obtained for $T\to0$. The exact expression for $T$ can be found in \cite{biref1}. Here we will only study the low energy limit $kl/m^2\ll1$, where there is no electron-positron pair creation and, as a consequence, $T$ is real. The induced ellipticity is in general accompanied by a rotation of the major axis, see e.g. \cite{DiPiazza:2006pr}. To lowest order, however, the ellipticity $\delta$ is simply given by the amplitude of $\epsilon'\langle E\rangle$ divided by the amplitude of $\epsilon\langle E\rangle$, with $\epsilon'\epsilon=0$, so $\delta=T_{\epsilon'\epsilon}$. Hence, the ellipticity induced on the probe field is given by the amplitude for a single (probe) photon to flip polarisation. The amplitude $T$ is in this regime given by a single lightfront time integral
\be\label{Tplanewave}
T_{\epsilon'\epsilon}=\frac{\alpha}{90\pi}\frac{kl}{m^4}\int\ud(kx)(-7\epsilon'\epsilon a'^2+3\epsilon'a'\epsilon a')=\frac{\alpha}{90\pi}\frac{1}{E_c^2}\int\!\frac{\ud kx}{kl}(7\epsilon'\epsilon(lf^2l)+3lf\epsilon' lf\epsilon) \;,
\ee
where only the second term in each expression remains if $\epsilon'\epsilon=0$. The coefficients in (\ref{Tplanewave}) allow us to deduce the low energy effective action \cite{biref1}; it is of course the Euler-Heisenberg action. The last form of (\ref{Tplanewave}) can be rewritten as an integral over the worldline of a probe photon $x_l^\mu=x_0^\mu+l^\mu nx/nl$, by changing variable from $kx$ to another, arbitrary lighfront time $nx$ \cite{biref2}. This allows us to generalise (\ref{Tplanewave}) by replacing the plane wave with an arbitrary background. In \cite{biref2} it is shown that, under certain conditions, this simple procedure actually gives the correct amplitude. In the next section we will investigate when it also leads to the correct ellipticity.

\section{Ellipticity}\label{Ellipticity-section}

We have just seen that, in the plane wave case, the ellipticity $\delta$ induced on the probe is given by the amplitude for a single probe photon to flip polarisation, which in turn can be written as an integral over the photon worldline. In this section we will show, without worrying to much about details, that, under certain conditions on the probe, the ellipticity induced by a general low energy background is again equal to the flip amplitude \cite{biref2} and to the wordline integral. As in \cite{biref1} we obtain $\delta$ from the expectation value of the electric field operator. However, this time we consider the low energy limit from the start, using the Euler-Heisenberg \cite{Euler:1935zz,Heisenberg:1935qt} effective action, which can be written without the dual tensor as \cite{Davila:2013wba}
\be\label{EHL}
\mathcal{L}_{EH}= \frac{1}{4} \text{tr}\, F^2 + \frac{\alpha}{90\pi}\frac{1}{E_S^2}\bigg[ \frac{7}{4} \text{tr}\, F^4 - \frac{5}{8}(\text{tr}\, F^2)^2\bigg]= \frac{1}{4} \text{tr}\, F^2+\mathcal{L}[F] \;.
\ee

In this section both the probe $P$ and the background $f$ are described by a coherent state $\ket{P,f}$. In the previous section the background was a plane wave and we chose coordinates so that it depended on $x^\LCp$. In this section, instead, $f$ is a general low energy background and we choose coordinates to make the expressions for the {\it probe} simple \footnote{This also differs from the coordinates used in \cite{biref2}.}. The probe field we have in mind is accurately described by a Gaussian beam (an approximate solution to Maxwell's equation) with focal spot radius $w_0$, pulse length $1/\Delta\omega$, wavevector $l_\mu=\omega(1,0,0,-1)$, and Rayleigh range $z_0=\omega w_0^2/2$. The field of the probe can be written $P_{\mu\nu}=l_{[\mu}\epsilon_{\nu]}P(x)$, where the pulse shape is given by
\be\label{Pulse_shape}
P(x)=\text{Re }P_0\zeta\exp-\Big(\frac{\Delta\omega^2}{4}{x^\LCm}^2+i\omega x^\LCm+\zeta\frac{x_\LCperp^2}{w_0^2}\Big) \;,
\ee
with $\zeta(z)=1/(1+iz/z_0)$. This is called the pulsed paraxial beam and the limit $\Delta\omega\to0$ is the paraxial beam \cite{Davis:1979zz,McDonald:1995}. The terms neglected are small when $\Delta\omega/\omega\ll1$ and $w_0/z_0\ll1$, which describe a field with a wavelength $\lambda=2\pi/\omega$ much shorter than the (focal) size, $\lambda\ll w_0\ll z_0$ and $\lambda\ll 1/\Delta\omega$. Indeed, at HIBEF \cite{HP}, for example, the width $w_0$ and length $1/\Delta\omega$ are measured in $\mu$m, $z_0\sim$mm and $\omega\sim10$keV, corresponding to a wavelength several orders of magnitude smaller than the size of the probe.

The expectation value of the total electric field is given by
\be\label{expval1}
\langle F_{\mu\nu}\rangle(x)=\bra{P,f}S^\dagger(t)F_{\mu\nu}(x)S(t)\ket{P,f},
\ee  
where $S(t)$ is the time evolution operator, and $F_{\mu\nu}$ is the electromagnetic field operator. Since $\ket{P,f}$ is coherent it can be transformed into a classical field by replacing $\ket{P,f}$ with the vacuum $\ket{0}$ and shifting the field operator (which is implicitly included in $S$) $F_{\mu\nu}(x)\to F_{\mu\nu}(x)+P_{\mu\nu}(x)+f_{\mu\nu}(x)$, where $P$ and $f$ are classical fields (this property of coherent states is described in \cite{Mandel-Wolf}). Then $\langle F\rangle=P+f+\delta F$, where birefringence effects are contained in the last term, which to lowest order in $\alpha$ is given by
\be
\delta F_{\mu\nu}(x)=i\int\limits^t\ud y\bra{0}[F_{\mu\nu}(x),\mathcal{L}[F+P+f](y)]\ket{0} \;.
\ee 

Note that up to this point the probe and the background are treated on equal footing. Under the assumption that the probe is weak compared to the background, we expand $\mathcal{L}[F+P+f]$ and keep the terms which are linear in $F$ and $P$ and quadratic in $f$. We then find
\be\label{deltaF}
\delta F_{\mu\nu}(x)=\int\ud y\;G_\text{ret}(x-y)\{\partial_{[\mu}\partial_{[\alpha} g_{\nu]\beta]}Pf^2\}(y) \;,
\ee
where $G_\text{ret}$ is the retarded Greens function and $\{\partial_\mu\partial_\alpha g_{\nu\beta}Pf^2\}$ is shorthand for the terms in $\mathcal{L}[\partial_\mu\partial_\alpha g_{\nu\beta}+P+f]$ which are quadratic in $f$ and linear in $P$ and $\partial_\mu\partial_\alpha g_{\nu\beta}$ (the derivatives act on $Pf^2$). We note in passing that (\ref{deltaF}) can also be obtained by solving the modified Maxwell's equation that follows from varying the effective action $S_{EH}$. The latter approach to birefringence has been used in \cite{DiPiazza:2006pr,King:2013zz,King:2012aw}.

Assuming that, apart from the electron mass, the frequency of the probe is the largest scale (cf. assumptions made in deriving the amplitude in \cite{biref2}), then $\partial_\mu\partial_\alpha\to-l_\mu l_\alpha$. This is certainly true at HIBEF \cite{HP}, where the background will have an optical frequency $\sim1$eV and a size measured in $\mu$m, and with probe parameters as above. Since we are interested in measuring the ellipticity of the probe a long time after interaction, and since the probe is centred at $\{x^\LCm,x^\LCperp\}=0$, we will evaluate the expectation value at $x^\LCp\sim 2t\to\infty$ and $\{x^\LCm,x^\LCperp\}\sim0$. In this region the Greens function simplifies
\be
G_\text{ret}(x-y)=\frac{1}{2\pi}\theta(x^0-y^0)\delta(x-y)^2\to\frac{1}{2\pi}\frac{1}{x^\LCp}\delta(y^\LCm-x^\LCm)
\ee 
and allows us to perform the $y^\LCm$ integral. Since the wavelength of the probe is small we can let $x^\LCm$ be small and still see many oscillations. In particular, we let $x^\LCm$ be small enough that we can neglect it in $\zeta(y)$ and $f$\footnote{Neglecting $x^\LCm$ in $f$ is not essential, we could keep it and have an $x^\LCm$ dependent $\delta$.}. Then the expectation value becomes
\be\label{epspE}
\epsilon'\langle E\rangle=\text{Re }\Big(-\epsilon'\epsilon-\frac{i}{2}\int\frac{\ud y^\LCp}{2\omega}\frac{\ud y^\LCperp}{\pi w_0^2}\zeta e^{-\zeta y_\LCperp^2/w_0^2}\{F_iF_of^2\}\Big)\frac{2i z_0}{x^\LCp}\omega P_0e^{-\frac{\Delta\omega^2}{4}{x^\LCm}^2-i\omega x^\LCm} \;,
\ee
where $\zeta=1/(1+iy^\LCp/2z_0)$, $F^i_{\mu\nu}=l_{[\mu}\epsilon_{\nu]}$ and $F^o_{\mu\nu}=l_{[\mu}\epsilon'_{\nu]}$. Notice the probe spreads out\footnote{In experiments a lens is used to recollimate the probe after interaction \cite{HP}.} when $x^\LCp\to\infty$; the field goes like $1/x^\LCp$, and $x^\LCperp$ has dropped out to lowest order in $1/x^\LCp$. In \cite{DiPiazza:2006pr} a similar expression was found for a different experimental setup.

(\ref{epspE}) is analogous to (\ref{delta-T-plane-wave}). In fact, performing the traces leads to
\be\label{trace}
-\frac{1}{2}\{F_iF_of^2\}=\frac{\alpha}{90\pi}\frac{1}{E_S^2}(7\epsilon'\epsilon lf^2l+3lf\epsilon' lf\epsilon) \,,
\ee
which is exactly the same vector structure as in (\ref{Tplanewave}). If we further assume that the probe is sufficiently narrow compared to the background, then we can neglect $y^\LCperp$ in $f$, perform the Gaussian $y^\LCperp$ integral and end up with a real integral over $y^\LCp$, which, as we will see in the next section, is again equal to the $\epsilon\to\epsilon'$ amplitude $T_{\epsilon'\epsilon}$. In other words, the factor in the round brackets of (\ref{epspE}) becomes $-\epsilon'\epsilon+iT_{\epsilon'\epsilon}$. The polarisation of the probe is again elliptical and to lowest order the ellipticity $\delta$ is equal to the ratio of the amplitudes of $\langle\epsilon'E\rangle$ and $\langle\epsilon E\rangle$, with $\epsilon'\epsilon=0$. All $x$ dependences, including $1/x^\LCp$, cancel in the ratio, and in terms of the worldline of the probe centre $y^\mu=l^\mu ny/nl$, with $ny=y^\LCp$, we find
\be\label{ellipticity}
\delta=T_{\epsilon'\epsilon}=\frac{\alpha}{30\pi}\frac{1}{E_S^2}\int\frac{\ud y^\LCp}{2\omega}lf\epsilon'lf\epsilon(y^\LCp,0,0)=\frac{1}{30\pi}\frac{1}{E_S^2}\int\frac{\ud ny}{nl}lf\epsilon'lf\epsilon(y^\mu) \;,
\ee
which is exactly what one finds by extrapolating the plane wave result. The fact that we are left with an integral over lightfront time means that the probe travels in a straight line without reflection, and effectively sees a plane wave background. The ellipticity is thus identical to the worldline expression for the polarisation flip amplitude found in \cite{biref2}, which we will discuss further in the next section. 

In upcoming experiments the probe might not be sufficiently narrow to allow us to completely neglect corrections. However, we can still find a relatively simple expression for $\delta$ if we assume that the background effectively restricts the $y^\LCp$ integral to scales small compared to the Rayleigh range $z_0$ of the probe\footnote{Note that, due to $\Delta\omega$ the pulse can still be short, which is something that will be relevant in the next section.}, in which case $\zeta\approx1$ in (\ref{epspE}). Two $E$-field components are again out of phase and the ellipticity is given by
\be\label{ellipticity_yperp}
\delta=\int\frac{\ud y^\LCperp}{\pi w_0^2}e^{-y_\LCperp^2/w_0^2}T[x_c^\mu]\;,
\ee
where the wordline (parameterised by $\phi$) is given by $x_c(\phi)=(x^\LCp_c,x^\LCm_c,x^\LCperp_c)=(\phi,0,y^\LCperp)$. (\ref{ellipticity_yperp}) takes into account the width of the probe by averaging the worldline amplitude over the transverse distribution of photons in the beam, and in the limit $w_0\to0$, it reduces to the worldline integral (\ref{ellipticity}).

\section{Probability of polarisation-flip}\label{Probability for polarisation-flip}

The derivation of the amplitude in \cite{biref2} was based in momentum space. This section gives an alternative derivation that facilitates comparison with the above derivation of the induced ellipticity and the expectation value of the number of flipped photons. Since we are still interested in the low energy weak field regime, we continue to work with the Euler-Heisenberg effective action (\ref{EHL}). Consider a probe photon, which initially has momentum $l_\mu$ and polarisation vector $\epsilon_\mu(l)$. After passing through a background $f$ there is a probability that it has flipped polarisation $\epsilon\to\epsilon'$, with $\epsilon'\epsilon=0$. In the parameter regime that we are interested in here, the photon momentum is approximately conserved, so we will be unable to see effects such as photon reflection~\cite{Gies:2013yxa}. The photon is described by the initial state
\be
\ket{\text{in}}=\tint{p}\;\psi\epsilon^\mu(p)a^\dagger_\mu(p)\ket{0} \;,
\ee
where $\psi(p)$ is a wavepacket peaked around $l_\mu$. The probability for polarisation flip is obtained using the Euler-Heisenberg action, leading to 
\be\label{Pstart}
\mathbb{P}(\epsilon\to\epsilon')=\tint{l}'\Big|\int\ud^4x\{\psi(x)F_o(x)f^2(x)\}\Big|^2 \;,
\ee
where $F^o_{\mu\nu}(x)=l'_{[\mu}\epsilon'_{\nu]}e^{il'x}$ and
\be
\psi_{\mu\nu}(x)=\tint{p}\;\psi(p)p_{[\mu}\epsilon_{\nu]}e^{-ipx} \;.
\ee

We immediately recognise the vector structure in (\ref{Pstart}) from the previous section, with $\psi_{\mu\nu}$ instead of the probe field $P_{\mu\nu}$. $\psi_{\mu\nu}$ resembles an EM field, but it need not be real and a single-photon state has zero expectation value of the EM field. Nevertheless, it solves Maxwell's equation $\partial_\mu\psi^{\mu\nu}=0$ and enters expectation values of quadratic observables in a way similar to an EM field, e.g. the energy momentum tensor $\bar{\psi}_{(\mu\tau}{\psi^\tau}_{\nu)}+...$ It might help to imagine $\psi$ having a shape similar to (\ref{Pulse_shape}). If the photon is localised, with respect to the background, in $x^\LCperp$ as well as in $x^\LCm$ (which would correspond to small $w_0$ and $1/\Delta\omega$), so that $\psi(x)f^2(x)\approx\psi(x)f^2(x^\LCp)$, then the $\{x^\LCm,x^\LCperp\}$-integrals yield a delta function\footnote{Similar simplification occurs, without assuming localisation, if the background only depends on one coordinate, e.g. lightfront time for a plane wave.} setting $p=l'$,
\be
\mathbb{P}=\tint{l'}|\psi(l')|^2\Big|\int\!\frac{\ud nx}{2nl'}\{F^iF^of^2\}\Big|^2 \;,
\ee
where $F^i_{\mu\nu}=l'_{[\mu}\epsilon_{\nu]}$ and $F^o_{\mu\nu}=l'_{[\mu}\epsilon'_{\nu]}$. If $\psi(l')$ is centred around $l$ with a width that is small compared to the characteristic energy of the photon, then using (\ref{trace}), we find
\be\label{final}
\mathbb{P}=\Big|\frac{\alpha}{30\pi}\frac{1}{E_c^2}\int\!\frac{\ud nx}{nl}lf\epsilon' lf\epsilon\Big|^2=|T|^2 \;,
\ee
which is precisely the worldline integral in \cite{biref2} and (\ref{ellipticity}). 

We note that to arrive at a worldline integral for $\delta$ (\ref{ellipticity}) we only had to assume localisation in the transverse directions $x^\LCperp$, while for the amplitude (\ref{final}) we have also assumed localisation in the longitudinal direction $x^\LCm$. The reason is that for $\delta$ we have the freedom to choose $x^\LCm$, whereas the flip probability is obtained by integrating over $x^\LCm$. Actually, from an experimental point of view, it might be more natural to consider the number of flipped photons rather than the ellipticity. The former can be obtained from the expectation value of the photon number operator with methods similar to those used here, and is closely related to the flip probability with the wavepacket replaced by the probe field. This is something that we will investigate further in a sequel paper.

\section{Discussion and conclusion}

We have complemented \cite{biref1,biref2} by providing an expression for the ellipticity $\delta$ induced on a probe passing through a general low energy background. Under certain conditions it reduces to the worldline integral expression for the polarisation flip amplitude $T$ found in \cite{biref2}. We have also presented a second derivation of $T$, which gives a slightly different perspective, being based in position space instead of momentum space. One of the assumptions leading to the wordline integral is that the probe should be narrow with respect to the background. Correction to this approximation for the ellipticity is given by (\ref{ellipticity_yperp}). It should be noted though, that in the interest of going further towards a more accurate description of upcoming experiments, one should instead of the ellipticity look at the expected number of flipped photons. As noted in the end of the last section, the expected number of flipped photons is closely related to the flip probability. This will be investigated elsewhere.

\begin{acknowledgements}
I am honored to receive a ``Gary McCartor Travel Award" and am grateful to ILCAC for this and for the opportunity to attend LC2014. I thank my collaborators in \cite{biref1,biref2}, especially A. Ilderton for a critical reading of the manuscript. 
\end{acknowledgements}

\end{document}